\documentclass[a4paper,10pt,twocolumn,hidelinks]{article}
\usepackage{amssymb,amsmath}                   % Mathematical Symbols
\usepackage{graphicx}                          % Figuresjor
\usepackage{tikz}
\usetikzlibrary{arrows,automata,shapes}
\usepackage[sort,numbers]{natbib}     % Bibliography
\usepackage[justification=centering]{caption}
\usepackage[utf8]{inputenc}
\usepackage{pst-pdf}
\usepackage{pst-node}
\usepackage{xcolor}
\usepackage{soul}

\usepackage{pdfpages}
\usepackage{lineno}
\begin{document}\title{\vspace{-2.8cm} Dynamical coupling during collective animal motion}
\author{T.O. Richardson$^{a,b,*}$, N. Perony$^{a,*}$, C.J. Tessone$^{a}$, \\ C.A.H. Bousquet$^{c}$, M.B. Manser$^{c}$, F. Schweitzer$^{a}$}

\date{\small
$^{*}$ These authors contributed equally to this work.\\
$^{a}$ Chair of Systems Design, ETH Zurich, Weinbergstrasse 56/58, 8092 Zurich, Switzerland\\
$^{b}$ Department of Ecology and Evolution, University of Lausanne, Biophore, 1015 Lausanne, Switzerland\\
$^{c}$ Institute of Evolutionary Biology and Environmental Studies, University of Zurich, Winterthurerstrasse 190, 8057 Zurich, Switzerland
}

\twocolumn[
\begin{@twocolumnfalse}
\maketitle

Major classification: Biological Sciences

\vspace{0.25cm}
Minor classification: Ecology

\vspace{0.25cm}

\vspace{0.25cm}
Corresponding author: Thomas Richardson, Office 3104, Department of Ecology and Evolution, University of Lausanne, Biophore, 1015 Lausanne, Switzerland. $+41\ 21\ 692\ 4181$. 
thomas.richardson@unil.ch

\vspace{0.25cm}
Author contributions: MBM \& CAHB designed and performed the data collection. NP \& CAHB constructed GPS collars for data collection. TOR \& NP analysed the data and wrote the paper. TOR, NP, CJT, MM \& FS provided analytical tools. FS \& MBM acquired funding, provided research facilities and provided support and advice to junior colleagues. All authors discussed the results and implications.
\end{@twocolumnfalse}
]

\newpage
\clearpage

%%%%%%%%%%%%%%%%%%% Main paper %%%%%%%%%%%%%%%%%%%%%%%%%%%%%%%%%%%%%%%%%%%%%%%%%%%%%%%%%%%%%%%%%%%%%%%%%%%%%%%%%%%%%%%%%%%%%%%

\begin{abstract}  %% Summary; 238 words. 
The measurement of information flows within moving animal groups has recently been a topic of considerable interest, and it has become clear that the individual(s) that drive collective movement may change over time, and that such individuals may not necessarily always lead from the front.
However, methods to quantify the influence of specific individuals on the behaviour of other group members and the direction of information flow in moving group, are lacking on the level of empirical studies and theoretical models.
Using high spatio-temporal resolution GPS trajectories of foraging meerkats, \emph{Suricata suricatta}, we provide an information-theoretic framework to identify dynamical coupling between animals independent of their relative spatial positions. Based on this identification, we then compare designations of individuals as either drivers or responders against designations provided by the relative spatial position. We find that not only does coupling occur both from the frontal to the trailing individuals and \emph{vice versa}, but also that the coupling direction is a non-linear function of the relative position.
This provides evidence for (i) intermittent fluctuation of the coupling strength and (ii) alternation in the coupling direction within foraging meerkat pairs.
The framework we introduce allows for a detailed description of the dynamical patterns of mutual influence between all pairs of individuals within moving animal groups. We argue that applying an information-theoretic perspective to the study of coordinated phenomena in animal groups will eventually help to understand cause and effect in collective behaviour.
\end{abstract}

% \section*{Significance statement}
% Nature is rife with examples of moving animal groups in which the collective seems to act as though it were a single organism. Clearly in such groups individual behaviours are tightly coupled. Nevertheless these groups often contain overly-influential `driver' individuals. Thus far, studies of collective motion have used simple linear correlations to identify leader-follower relationships, despite the fact that even coupled systems do not necessarily produce linear correlations. 
% 
% Here we present an entropic framework that allows such action-reaction relationships to be reformulated in terms of fundamental directional information flows, and fully captures nonlinear dependencies. We apply this technique to spatial trajectories obtained from pairs of GPS-collared meerkats and reveal that driver and responder roles dynamically vary within a pair.

%%%%%%%%%%%%%%%%%%%%%%%%%%%%%%%%%%%%%%%%%%%%%%%%%%%%%%%%%%%%%%%%%%%%%%%%%%%%%%%%%%%%%%%%%%%%%%%%%%%%%%%%%%%%%%%%%%%
\section*{Introduction}                  
% Describe the 'phenomenon'
To date, most investigations of the organisation of collective movement have focused on cases of coherent group movement, in which local speed or velocity differences between individual and group motion are rather minor. This may be the typical case in bird flocks or fish shoals, but the same may not be so for moving terrestrial groups. In such groups \citep{bazazi2012}, individual movement often consists of periods of rapid displacement during which the animal moves in concert with the group, and is intermittently punctuated by near-stationary phases during which it may engage in local area-restricted search, for example after encountering a food-resource patch. In such groups, motion at the individual and  group levels may not be so strongly synchronised as is observed in flying or swimming groups; at any moment, a sizeable fraction of the group may perform area-restricted movement, whilst the rest continue onwards.
% State the problem
The leading edge of a moving animal group may create one or more dynamical fronts whose direction of movement is subject to sudden changes, whose shape is subject to abrupt deformations, and whose membership is in a constant state of flux. How then to reliably identify the influential individuals? 

% Recap previous experimental results & describe the challenge; formal identification of leaders.
Studies now abound on the role of influential individuals (such as leaders) in collective behaviour. Theoretical progress has been rapid \citep{gregoire2003,rands2003,rands2008,gregoire2004,couzin2005,conradt2009}, and often outstrips empirical progress. Nevertheless, in experimental settings it has been shown that variation in nutritional state \citep{sueur2010,mcclure2011,furrer2012}, in indicators of mate quality \citep{toth2011}, in temperamental \citep{harcourt2009} or personality \citep{nakayama2012,nakayama2013} traits such as boldness, in relatedness between leader-follower pairs \citep{lewis2013}, or in differential knowledge of food \citep{reebs2000,pillot2010,bousquet2011b} and nesting locations \citep{stroeymeyt2011} can all give rise to the emergence of leading or following roles. Such studies often operationally define leaders as those individuals at the front of the group, although recently it has been put into question whether these are indeed the key individuals driving group movement \citep{
bumann1993,nagy2010,sueur2009,sueur2010b,katz2011}. Consequently, a major challenge remains in formally measuring the strength of influence that different individuals within the group exert upon one another, as only with a complete categorisation of all such links will it be possible to infer the leaders.
In what follows we term this influence `coupling' in order to  distinguish its low-level nature from the classical high-order definitions of leadership as a stable individual characteristic, associated with dominance rank, and issues of intentionality or control \citep{king2009}.

% set out the paper structure
In this paper we will argue that existing methods for measuring the influence within pairs of individuals -- namely, pairwise correlations -- do not fully capture the variety of functional dependencies that such coupling can take. We then present a more comprehensive entropic measure which explicitly quantifies the dependency in terms of information flow, rather than measuring correlations because such correlations are not always guaranteed to be present when there are underlying information flows. Using the simplest possible dataset -- pairs of simultaneously recorded individual spatial trajectories -- we demonstrate the basis of the approach. Finally we outline the limitations of this analysis, and advocate the extension of the approach to measuring the coupling between all unique pairs, from which a formal identification of the leader(s) will be possible.
However, we first describe how coupling relates to `leadership' as it is currently defined in the context of animal behaviour.

% Discuss higher-order definitions of leadership.
Classically, a leader in an animal group is an individual with some unique skill or behavioural characteristic which through its actions controls or directs the group's behaviour. In this everyday sense, the quality of leadership is some characteristic that is consistent over at least an intermediate time-period. This definition has been adopted by psychologists \citep{Smith1998}, ethologists and behavioural biologists \citep{dumont2005}. Essentially, a leader is a key individual whose impact on the collective behaviour is greater than that of others \citep{king2009}. 
Under this definition, the maximum number of individuals that may be ascribed the label of `leader' is undefined, as is the related question of whether the remaining non-leaders should all be equally responsive to the actions of the leader(s). So for clarity, here we consider the simplest case termed `personal leadership' \citep{leca2003}; a \emph{single} individual whose influence upon \emph{all} the other group members is much greater than that of any other individual.
Classical personal leadership often occurs in hierarchically-organised societies, in which a single dominant individual can control the actions of other subordinate group members \citep{squires1975,peterson2002}. In these cases, it is long-term individual traits that determine dominance, and hence status as the group leader \citep{kurvers2009}. 
%% Introduce sequential distributed leadership -- almost all in primates. 
However, the possession of a hierarchical social organisation does not mean that leadership is necessarily centralised upon the most dominant individual in all contexts. For example macaques \citep{thierry2007,Sueur2008,sueur2009,sueur2010b}, lemurs \citep{jacobs2008} and capuchins \citep{leca2003} possess dominance hierarchies, but in all cases the initiation of group movement is not associated with dominance, but instead sequentially alternates among all the group members. In this scenario, an individual is only a leader during the short period during which it influences group movement, so individual leadership is temporary and distributed.

%% Pairwise coupling is the basis of *both* higher-order leadership and sequential distributed leadership
Given the above, it seems clear that the fundamental feature common to both personal (centralised) and sequential (distributed) leadership is that the actions of the leader causally impact upon those of the other group members through a set of dynamical dependencies, known as `drive-response' coupling \citep{rosenblum2001,paluvs2007,vejmelka2008}. In the case of animal groups that have personal or sequential leadership, the `coupling direction' is by definition from the leader(s) to the follower(s). 
%% Note that there must be coupling between non-leader animals in groups with a classic personal leader
However, in many such cases at any given moment a sizeable proportion of the group may not have a potential direct link (e.g. line-of-sight) with the leader. Then, coupling between non-leader individuals is still necessary in order (i) for the influence of the central leader to indirectly propagate, and (ii) to maintain group cohesion. Indeed, in any animal group possessing a centralised leader one can measure the coupling between pairs of individuals, without knowing how far down the causal chain one is observing. 
Therefore, it is necessary to recognise that such coupling mechanisms must also be in operation in situations in which at any given moment there are numerous individuals that simultaneously but fleetingly exert an influence at the group level, but which have neither personal nor sequential leadership, for example in large yet cohesive flocks \citep{cavagna2010a}, swarms \citep{janson2005}, shoals \citep{reebs2000} and herds \citep{prins1995}.

%% Our claim!
Our central claim is that dynamical coupling mechanisms operating between pairs are the fundamental components common to not only all forms of leadership, but also all cases where a group has no leaders \emph{per se} but still exhibits cohesion \citep{gregoire2003,buhl2006}. This assertion is perhaps best evidenced by considering a cohesive moving animal group in which a single centralised leader is present, but where the individuals do not possess a global overview. 

%% coupling is the basis of higher-order leadership
The primary challenge in King's \citep{king2009} definition of  leaders as those individuals that disproportionately influence the other group members, is in quantifying the magnitude of the causal links from the leader to each of the group members. However, as the fundamental mechanism underlying all forms of leadership is pairwise dependencies (coupling), the same challenge exists when measuring the coupling within a pair of non-leader individuals, although then the challenge is more complex as the direction of the link (the coupling direction) is not given \emph{a priori}.

%% Lay it out...
As the coupling within a pair is only a sample from a large set of simultaneously-occurring pairwise interactions, it represents a window into the inner dynamics of the group that will likely include many causal interactions not necessarily associated with the central leader, for example, those associated with play, sexual competition, foraging competition, group cohesion, predator detection and evasion.

% introduce previous studies that DID explicitly study causation by examining time-lagged correlations
By analysing the time ordering in the movements (spatial trajectories) in animal pairs, several studies have made important methodological progress towards quantitative inference of pairwise coupling \citep{nagy2010,herbertread2011,ward2011,katz2011}. Such studies tacitly subscribe to the long-standing convention of describing some event, A, as causal to another, B, only if A precedes B, and if knowing A increases the prediction ability about the future of B \citep{wiener1956,granger1969,suppes1970}, rather than the more rigorous (but often experimentally intractable) definition that requires demonstration of cause and effect through experimental intervention. Although experimental manipulations of coupling are possible, for example through insertion of robotic `dummy' individuals into a moving group \citep{michelsen1992,faria2010,tarcai2011,marras2012}, none exists for field experimental systems such as cohesively foraging terrestrial species. Hence, here we are constrained to methods of inferring 
leadership non-manipulatively - typically from the time ordering of paired time series.
Such methods include Hellinger distance \citep{Kampichler2012}, Granger causality \citep{granger1969}, and perhaps more simply, lagged correlation analyses \citep{knox1981}. 

In the context of collective animal behaviour the lagged correlation analysis consists of measuring the correlation in the orientations or displacements within pairs of moving animals. This gives the coupling strength, and the behaviour of such correlations as a function of the time-delay gives the coupling direction, and hence, which -- if any -- leads \citep{nagy2010,herbertread2011,ward2011,katz2011,bode2011a}. However, to establish which individual in a pair drives (and hence which responds), these studies used the complete time-aggregated trajectories to establish the direction of causality over the entire observation period. Here, rather than using the complete trajectories to assign a fixed label to the relationship between a given pair,  we instead investigate the dynamics of the driver-responder relationship through the use of sliding windows to split the trajectories into a sequence of overlapping trajectory subsections (see Supplemental Information).
%% Make sure that the Supplemental Information we refer to is really a Supplemental Information and not a Supplementary Information!

% need a caveat to distinguish high-frequency within-bout leadership alternation in meerkats, from low-res between-bout alternation in primates.
In primate groups, the identity of the initiators of bouts of collective movement, such as the initiation of group departure from a resting site, may change from bout to bout \citep{stueckle2008, sueur2011}. However, the behavioural context we study here is not that of the initiation of group movement: instead, the trajectories we study comprise single coordinated foraging bouts. A further difference between the current study and those that focused upon the initiation of collective movement is the temporal resolution of the data; the high recording frequency of the trajectories studied means that a more natural comparison is with studies of tracking at a high temporal resolution. Such data allow for a fine-grained description of the dynamics of influence between pairs of individuals, specifically through the characterisation of the direction and magnitude of this influence. The uninterrupted measurement of individual position makes it possible to consider this influence from a near-continuous perspective 
rather than during discrete events, as has often been the case in past studies \citep{stueckle2008,sueur2011}. 

%\hl{For this reason, in the present paper we define leadership as a time-varying behavioural influence between two individuals, which is a dynamical, individual-oriented extension of} \citet{king2009}'s \hl{definition.}
% null hypothesis #1
The absence of empirical data on time dependence in pairwise drive-response relationships naturally leads to our first null hypothesis; that within a single bout of group foraging, the causal relationship between the actions of pairs of individuals is, if not of fixed magnitude, then at least constant in direction.
% Introduce the comparison to the classical definition of leadership via spatial position
In many experimental studies of moving animal groups the operational definition of an individual as a leader was \citep{squires1975,bumann1993} -- and still often is \citep{burns2012,van2012} -- defined by its occupancy of a position at the leading edge of the moving group. Theoretical approaches to leadership have found that when the model input includes between-individual variation in the preference to orientate towards some goal, such as an attractive location \citep{couzin2005}, or to move according to an external gradient \citep{guttal2010}, then the individuals with prior knowledge or the stronger preference occupy the frontal positions.

% null hypothesis #2
Our second null hypothesis is the following: individuals towards the leading edge of the moving group influence those in other positions to a much greater extent than do those at the trailing edge.
This proposition is motivated by the argument that, whilst in practice the individuals that most influence the group's direction of motion are often found at the front \citep{bumann1993,franks2006}, it is not clear \emph{a priori} that leadership by relative spatial position and leadership by copying ({\em sensu} King \citep{king2009}) need be necessarily correlated. In principle, an individual whose actions strongly influence others in the group may occupy non-frontal positions.

To disentangle the potentially confounding correlation between spatial position and causation, we first apply an information-theoretic framework for measuring coupling in moving animal groups. Then, using paired high-frequency GPS trajectory data from a model empirical system \citep{bousquet2011}, collective foraging in meerkats, we demonstrate the applicability of the approach to the extraction of drive-response relationships from noisy field data.
% TOM: MOVED UP FROM THE RESULTS
Interestingly, similar techniques have been used to track waves of information diffusion through cortical neural networks \citep{garofalo2009,wibral2011} and simulated animal groups \citep{wang2012}. However, to the best of our knowledge, this study is the first to compare entropic driver-responder classifications with position-based classifications of leader-follower roles, and the first to apply the approach to empirical data on collective behaviour on animal groups. 
% Note that, in principle one could consider all unique pairs of animals within the group (for which simultaneously-recorded trajectory data are available), here we present an analysis of single pairs. 

%%%%%%%%%%%%%%%%%%%%%%%%%%%%%%%%%%%%%%%%%%%%%%%%%%%%%%%%%%%%%%%%%%%%%%%%%%%%%%%%%%%%%%%%%%%%%%%%%%%%%%%%%%%%%%%%%%%
\section*{Results}

%% SHORTEN THIS SECTION !!
\subsection*{Establishing the coupling strength and direction using an entropic approach}
Our framework for extraction of coupling rests on four steps. 
% TOM: REMOVED THIS PREFACE AS EACH PARA INTRODUCES ITSELF ANYWAY....
%(i) selection of observed paired animal trajectories, (ii) use of surrogate data techniques to construct ensembles of pairs of synthetic trajectories which by construction contain no coupling, (iii) use of an entropic measure \citep{shannon1948,mann2011} to find at each time step the coupling direction between the two observed animal trajectories as well as between the artificial trajectory pairs, and (iv) extraction of the statistical significance of the observed coupling by comparison with that measured in the surrogate trajectories.
In the first step of the framework, a pair of animals, $A,B$, is selected from the group, and their respective time-stamped trajectories are converted into a paired sequence of displacements, $A_t,B_t$, that is into time series. These displacements can be either in one axis of motion only, or within the plane itself (see the Observables subsection in the Methods).

In the second step, we generate pairs of statistically realistic null-model trajectories which start and end at the same locations as observed, but which are otherwise statistically independent of one another. To do so, we turn to a class of constrained randomisations termed surrogate data techniques to generate synthetic trajectories \citep{prichard1994,schreiber2000,pereda2005}. The primary advantage of data surrogates over standard shuffling methods is that they conserve within-trajectory correlations -- such as those within the sequence of $(\delta x,\delta y)$ displacements that make up each trajectory -- and so produce more realistic trajectories (Fig.~\ref{fig:METHOD}, see Supplemental Information for further details).

In the third step, the net coupling direction and the coupling magnitude between $A$ and $B$ are obtained by comparing the directional information flows between the paired time series, $A \rightarrow B$, and $B \rightarrow A$, from which we assign a causal direction to their interactions; if the aggregate information flow is predominantly $A \rightarrow B$, then the movement of $B$ is driven by the prior actions of $A$, hence $A$ influences $B$.
To quantify directional information flows we calculated the Mutual Information and the Conditional Mutual Information (henceforth, MI \& CMI) between the two trajectories \citep{paluvs2007,vejmelka2008}. 

% Describe the metric in more detail -- required because results section follows immediately.
Although inference of the coupling direction can be made using time-lagged correlations, such as Pearson's $r$, such approaches inherently assume a linear drive-response relation. As we do not wish to make any \textit{a priori} assumptions regarding the functional form of the dependency between the displacements of the putative driver and those of the responder, we use entropic measures \citep{shannon1948,mann2011} of the dependence between paired (bivariate) distributions which, unlike correlations, do not make assumptions about the functional form (e.g. linearity) of the dependence (Fig.~S2).
% TOM: MOVED THIS TO THE  SI (LATEX L.90) ALONG WITH FIG. 1.
%If the relationship between the purported driver and response variables is thought to be non-linear, then non-parametric correlation metrics (e.g. Spearman's $\rho$) are often preferred due to their ability to capture non-linear relationships. However, if for each value of the predictor there exists more than one response, even non-parametric correlations fail \hl{(Fig.~S2)}. %\ref{fig:CORR_V_MI}) 
In contrast with parametric and non-parametric measures of correlation, information-theoretic measures of dependence unveil relationships not between paired values of predictor and response, but instead compare the information held \emph{within} their joint (bivariate) distribution that cannot be derived from their marginal distributions.

In short, the MI quantifies how much the current state of a given time series, $A_t$ (the putative driver) reduces the uncertainty in the future state of a partner, $B_{t+lag}$ (the putative follower), and so provides an estimate of the coupling strength between paired time series. This coupling is denoted $I(A_t;B_{t+lag})$. However, as the MI from the present $A$ to the future $B$ is symmetrical to MI from the future B to the present A, $I(A_t;B_{t+lag})=I(B_{t+lag};A_t)$, the MI contains no inherent direction information. 
For a directional measure of information transmission from one time series to another, or in other words, how much the first influences the second, we use the CMI, $I(A_t;B_{t+lag}|B_t)$.  Like the MI, the CMI measures information transmission from $A_t$ to $B_{t+lag}$, whilst also including a conditional term describing the current actions of the putative follower, $B_t$. This removes any self-influence of the putative follower upon its future actions, with the residual measure quantifying the influence of the current actions of the putative driver upon the putative follower. This process is illustrated in Fig.~\ref{fig:TS_3D}. For simplicity, the CMI from $A_t$ to $B_{t+lag}$ will be written $I(A \rightarrow B)$. 
An important outcome of the conditionality term in the CMI is that, unlike the MI, the information flow is no longer symmetrical; $I(A \rightarrow B) \neq I(B \rightarrow A)$. Hence, by comparing the asymmetry of the two directional CMI flows within a pair with the summed flow in both directions, thus, 
\begin{eqnarray*}
  D(A \rightarrow B)=\frac  {I(A \rightarrow B)  - I(B \rightarrow A) } 
                            {I(A \rightarrow B)  + I(B \rightarrow A) }
\end{eqnarray*}
we obtain the net normalised information flow, termed directionality (Rosenblum \& Pikovsky, 2001). This measure thus allows the inference of the causal role of `drivers' by measuring the degree to which their movements influence the future movements of others (`responders'), rather than through their \emph{relative} spatial position within the moving group.
%TOM: MOVED THIS UP TO THE END OF THE INTRO
%Interestingly, similar techniques have been used to track waves of information diffusion through cortical neural networks \citep{garofalo2009,wibral2011} and simulated animal groups \citep{wang2012}. However, to the best of our knowledge, this study is the first to compare entropic driver-responder classifications with position-based classifications of leader-follower roles, and the first to apply the approach to empirical data on collective behaviour on animal groups. 
% TOM: INCORPORATED THIS INTO THE SENTENCE L155
%By explicitly inferring causation through the extraction of the aggregate directionality of information flow, the CMI framework allows the identification of drivers not by their \emph{relative} spatial position within the moving group, but by the degree to which their movements influence the future movements of others.

Lastly, in the final stage of the framework the observed and expected directionalities (from the ensembles of surrogate trajectories) are statistically compared using two-tailed randomisation tests. These results are presented in the next section.

\subsection*{Coupling is intermittent and reversible} %% Re-state hypothesis #1 & how to reject it
By extracting a near-continuous and signed (rather than binary) dynamical coupling metric for pairs, and in combination with the comparison to the `null' trajectory pairs generated by the data surrogates procedures, we are able to test the first null hypothesis; that within a single foraging bout, pairwise coupling does not alternate. This hypothesis would be rejected in one of two cases, (i) if only one individual in the pair exhibited periods of statistically significant coupling which were interspersed by periods of no coupling, and (ii) if statistically significant coupling alternates from one individual to the other. 
%% Lead-up to the result
We used a two-tailed randomisation test to compare the observed directionality at each time step with the distribution of directionalities obtained from the surrogate pairs, to establish whether the observed directionality represented a statistically significant coupling event.
%% The result
For each session we then found the overall proportion of time steps in which statistically significant coupling was observed (Table~\ref{tab:PERCENT_SIGNIF}). As for all sizes of sliding window (except the smallest), this proportion was above chance levels (5\%), the null hypothesis of time-constant coupling between the members of the pair could be rejected.
%% One-sentence conclusion
In sum, for pairs of foraging meerkats, the causal drive-response relationship was neither of fixed magnitude, nor constant in direction (Fig.~\ref{fig:DIRECTIONALITY}). In other words, both the strength of influence the driver exerts over the responder and the identity of the driver itself vary over time.

\subsection*{Rapid decay of information flow with increasing time delay}
%% Describe information-flow verses time-delay
We observe a fairly rapid decay in the MI as a function of the time lag, which indicates the absence of long term correlations between the pair; all evidence of coupling is lost after a delay greater than about 20 seconds. However, the relationship is subject to considerable noise (inset of Fig.~\ref{fig:FLOW_VS_FB_DIST}a). For the reasons described at the beginning of this section, the MI alone is insufficient for a complete description of the net influence of one animal upon the other. Hence to measure the net information flow of each animal upon its partner (excluding the self-influence of each recipient upon itself), we measured the CMI. As this is not symmetrical, $I(A \rightarrow B)\neq I(B \rightarrow A)$, and as here we are only interested in examining how the total information exchange varies according to the time-delay between the cause and effect, we sum the two directional CMIs (Fig.~\ref{fig:FLOW_VS_FB_DIST}a), which shows a much cleaner decay. 
% add once-sentence take-home message
The decay of the information transmission with increasing time-lag indicates that a foraging meerkat responding to another individual may react to the movements of that driver up to about 15 seconds in the past, but that movements beyond this time horizon have no influence, and hence are probably not remembered.

\subsection*{Driving from the rear}

%%Re-introduce what we did & how to test it
We now turn to our second hypothesis; that the predominant direction of information flow is from the individual at the front to the one in the rear \citep{bumann1993,nagy2010,katz2011}. 
%% COPIED FROM ABOVE & slightly reworded
Were this the case, we should find that an individual towards the front of the moving group should transfer information to those behind, and relative longitudinal position (LP) alone would be the sole determinant of the coupling direction. Then one would expect that as animal A moves from rear to frontal positions (LP distance $<$ 0 to LP distance $>$ 0), $D(A \rightarrow B)$ also shows a monotonic increase from negative (responding) to positive (driving).
%% Re-state hypothesis #2 & how to reject it
By contrasting the driver-responder designations given by the relative spatial position with those derived from the causative relationships, we are able to explicitly measure the information each animal transmits to the others in the group, as a function of its relative spatial position. 
% THIS IS REDUNDANT, GIVEN L191
%For example, if the actions of those individuals at the front are more influential than those in other positions, then the net direction of information flow (as captured by the directionality) will point from those individuals occupying front positions to those individuals occupying other positions. Thus we would expect that as animal A moves from rear to frontal positions (LP distance $<$ 0 to LP distance $>$ 0), $D(A \rightarrow B)$ also shows a monotonic increase from negative (responding) to positive (driving).
%Such a mechanism would manifest as a positive relationship between the relative longitudinal distance and the directionality. 
For each pair we calculate the mean directionality across the entire trajectory, and use the sign of this average to identify one member of the pair as the overall driver, $L$. By implication we define the other as the overall responder, $F$ (here we use L and F in analogy with driver and responder to avoid potential confusion of the driver with the directionality, D). The second null hypothesis -- that the directionality $D(L \rightarrow F)$ is greatest when the driver is at the leading edge of the moving group -- could then be tested.
%% describe the results
The observed relationship of $D(L \rightarrow F)$ as a function of the relative longitudinal position of $L$ was neither positive nor linear (Fig.~\ref{fig:FLOW_VS_FB_DIST}b).
%% Describe the lines we draw through the plots.
To characterise these non-linear relationships, we used penalised splines because of their flexibility, and because they provide a data-driven non-parametric estimate of the relationship, without making any prior assumptions about its functional form \citep{ruppert2003}. 
Although the spline fits are provided mainly as a guide to the eye, and to highlight discontinuities in the information flow, they did provide statistically reasonable fits; for all plots the smoothed terms were significant at the p$<$0.005 level, and typically explained over half of the variance (mean adjusted R-squared=0.58).
The spline fits highlight two peaks in the directionality; one when the driver was at the front -- as expected -- but also a peak at the rear. The peak at positive LP values shows that, indeed, the individual in the front of the pair does influence the one in the rear, however, this fact alone does not allow us to confirm the second null hypothesis. Rather, the peak at negative LP values indicates that information may also propagate from the rear to the front, which leads to a rejection of the second null hypothesis.
%% Discuss the reversal of leadership roles at the extreme LP values; at distances above a threshold, attraction operates, which could promote group cohesion. 
Further, we note that at extreme values of the LP, the directionality rapidly switches from positive to negative. More specifically, an individual that is 7.5-10 metres in front of the pair's centre (that is 15-20 metres ahead of the other individual) strongly influences the trailing partner, yet if it gets any further ahead the roles swap, and it instead becomes reactive to the movements of the trailing individual. The converse is true for a trailing individual that falls too far behind its pair partner. 
%% Could move to discussion??
%This sensitivity could be interpreted as evidence for the presence of a tendency to revert to the mean group position beyond some threshold distance.
%% Compare spline fits to the baseline information flow in the surrogates
To ascertain whether the observed peaks in $D(L \rightarrow F)$ are meaningful, we performed the same driver identification process as described above but for each surrogate trajectory pair. We thus operationally define one individual in the surrogate pair to be the driver, $L_{surr}$, and so obtain the expected directionality of the individual that -- in the absence of any coupling -- \emph{appears} to drive, namely, $D(L_{surr} \rightarrow F_{surr})$, which can then be directly compared to $D(L \rightarrow F)$. We observe that (i) the observed peaks are far outside the value of their surrogate counterparts, and (ii) between -5~m and +5~m from the pair's centre, $D(L \rightarrow F)$ is typically not significantly different from $D(L_{surr} \rightarrow F_{surr})$, meaning that there is little to no net information flow within this zone. Hence we accept that the influence an individual exerts upon its conspecific strongly depends upon its relative spatial position within the pair (Fig.~\ref{fig:FLOW_VS_FB_
DIST}b).

%%%%%%%%%%%%%%%%%%%%%%%%%%%%%%%%%%%%%%%%%%%%%%%%%%%%%%%%%%%%%%%%%%%%%%%%%%%%%%%%%%%%%%%%%%
\section*{Discussion}

% Discuss coupling fluctuation; This para is a bit dry...
We have shown the existence of dynamical alternation of pairwise driver-responder roles within moving animal pairs, so we cannot reject the null hypothesis that within-bout pairwise drive-response relationships are fixed. 
The directionality time series contain periods of significant deviation from the expected value of zero, and visual inspection indicates that the periods in which either one or the other animal showed statistically significant coupling are somewhat clustered. Indeed, the temporal statistics of the directionality (see Supplemental Information) show the presence of non-trivial (positive) temporal correlations, which indicate that the coupling relationship does exhibit non-random temporal persistence. 
% Discuss the link between the dependence of influence upon spatial position, and the directionality fluctuation.
This fluctuation in both the identity of the driver and the strength of its influence upon the other group member should be viewed within the context of  the relative spatial position. Within a single foraging bout, an individual meerkat may occupy many different positions within the moving group, hence for any given position there is a considerable turnover of individuals. It is then natural that -- given the observed relationship between position and directionality -- its influence upon the other group members should wax and wane.
%% Discuss what the non-linear functional form of directionality ~ LP dist could mean.
The non-linear form of the relationship between relative spatial position and directionality may reflect the existence of an underlying set of ``rules of thumb'' underlying individual (and therefore also collective) movement \citep{couzin2002,rands2011}. For example, the front and rear directionality peaks represent positions (at the leading or trailing edge of the moving group, respectively), at which the motion of a given individual has the greatest causative influence upon the other member of the pair. Such positions might represent zones within the moving group within which biologically-relevant events occur -- such as discovery of new food items (front) or predator attacks (back). Hence, these are zones to which individuals may be dynamically attracted or repelled depending upon the movements of their conspecifics or the occurrence of heterospecifics. Similarly, the switch from being highly influential to becoming highly reactive (positive to negative directionality) that occurs beyond these peaks may 
be related to an individual tendency to reorient towards the group centre beyond some threshold distance. This phenomenon can be seen as the signature of some underlying interaction mechanism that functions to maintain group cohesion.

% Applicability to complex systems in which only a few degrees of freedom are accessible / use of various data sources in addition to movement logs
Clearly, the social activities of a group-living animal occur within a space consisting of much more than one degree of freedom. Yet the CMI-based approach was able to identify drive-response behaviours solely through examination of paired but univariate time series, namely the sequence of displacements within only a single axis of motion (Table~\ref{tab:PERCENT_SIGNIF}). 
In systems with many interacting degrees of freedom, only a few may be experimentally accessible. In such cases, investigators may acquire a reasonable picture of the overall system dynamics by measuring the information flow on one of those degrees of freedom -- something that promises to simplify future investigations (see also \cite{wicks2007}). Additionally, the time series used for analysis could in principle come from any source, and one could readily examine the information transmission between various data sources, for example the influence of animal vocalisations on acceleration measurements.
%% slag off other studies that claim to measure information transmission; could go in intro or discussion.
The overwhelming majority of studies on information transmission in animal groups, have not measured the information transmission \emph{per se}, but instead measured the dissemination of some proxy, typically the alignment correlation \citep{radakov1973,buhl2006,cavagna2010a,cavagna2010b,nagy2010,katz2011,herbertread2011,ward2011,bode2011a,bialek2012}. Although the presence of such correlations do indeed suggest that information has been transmitted (and hence that claims of coupling based on measures of correlation are legitimate), the presence of information transmission does not always produce a correlation \citep{sugihara2012}. This means that studies employing classical correlation-based methods may suffer from a bias towards the refutation of information-transmission hypotheses, that is, an increased type-II error rate. Future studies should consider the use of information-theoretic measures to maximise the chances of capturing cause and effect in animal groups.

%% caveats
It is important to bear in mind the caveat that the inferences drawn above are based on measuring the coupling between pairs of individuals within much larger groups. Whilst the pairwise information plots do show strong evidence for directional coupling -- that is for the presence of driver and responder roles within pairs -- it could also be that other individuals within the group (i.e. untagged meerkats, about which we had no GPS data), or stimuli external to the group (e.g. heterospecific warning calls), were actually driving one or both of the tagged individuals. Nevertheless, interdependencies between individuals far down the group movement chain are still to be expected \citep{sueur2009}, and what we describe as direct coupling is indeed the sum of both direct and indirect influences. In the present case, it is likely that direct influences prevail, since we observed only dominant pairs (see Methods) and direct interactions within the dominant pair are common (e.g. in the case of mate guarding \citep{
spong2008}). Importantly, there are factors affecting group coordination in meerkats that we did not observe, such as interindividual vocalisations \citep{townsend2011,bousquet2011}. Whilst the discrete and sporadic nature of vocalisations makes them challenging to study in combination with continuous movement data, future work will explore the link between these two dimensions of collective behaviour in meerkat groups.

%% overview & future applications of the framework
It is also worth noting that whilst the directionality does capture the asymmetry of information flow between a given pair of animals, it does not imply anything about the routes by which causative influence propagates. For example, it is possible that, rather than representing direct action-reaction interactions between a leader and a follower, many of the statistically significant interactions we detect actually represent unseen and indirect `domino' sequences, that is, chains of intermediaries through which the initial actions of the leader propagate. More explicitly, the effect of sampling single pairwise interactions within larger groups, combined with the presence of possibly indirect causative chains, is that rather than representing a direct causative relationship, from a leader, $A$, to a follower, $B$, (or $B$ to $C$), we instead sample two -- potentially downstream -- locations within a causative chain. For example, given the chain $A \rightarrow B \rightarrow C \rightarrow D$, we may sometimes 
measure $B \rightarrow D$. A natural extension of the current approach would therefore be to measure the directionality between all unique pairs of individuals in the group, and to then treat these time series as directed and weighted links in a time-explicit social network \citep{nagy2010}. Indeed, in cortical neural networks, measures of joint and conditional entropy have proven to be excellent alternatives to classical correlation-based methods for mapping the physical topology of the synaptic connections \citep{garofalo2009}. Only such a network-based approach can provide a complete mapping of all the pairwise relationships in the group, and only then will it be possible to define the social organisation as characterised by centralised or distributed leadership.

%% Perspective on the framework & applications to other systems
Whilst techniques for determining the directionality of causation based on time-lagged cross-correlations have a strong tradition in neurobiology \citep{knox1981} and economics \citep{granger1969,podobnik2010}, their use in the behavioural sciences is comparatively infrequent, mainly because in behavioural research hypothesis testing is typically undertaken using controlled experimental manipulations. Using the general framework described herein, one could for example in pilot studies generate working hypotheses that may then be experimentally tested. This work should be seen not as an alternative, but as a supplemental approach to the gold standard of experimental manipulation. We have provided the basic elements for a more complete quantification of pairwise dependencies in groups, which we hope will facilitate mapping of the dynamical patterns of connectivity within animal societies.

%%%%%%%%%%%%%%%%%%%%%%%%%%%%%%%%%%%%%%%%%%%%%%%%%%%%%%%%%%%%%%%%%%%%%%%%%%%%%%%%%%%%%%%%%%%%%%%%%%%%%%%%%%%%%%%%%%%
\section*{Methods} 
\subsection*{Study system}
Individual movement data were acquired from a wild population of meerkats at the Kalahari Meerkat Project, South Africa (26$\,^{\circ}$58'S, 21$\,^{\circ}$49'E). For a detailed description of the study site see \citep{clutton2001}. The GPS collars were attached on a day at least one week prior to data collection, during which the individuals were then caught and anaesthetised according to the protocol described by Jordan et al. \citep{jordan2007}. A single GPS recording session involved attaching a GPS tag to the pre-fitted collars worn by the dominant male and female in a group, as dominant individuals have been shown to have the greatest influence on group movement \cite{perony2013}. These individuals were then allowed to forage as usual. The six GPS sessions were acquired from five meerkat groups (Aztecs, DrieDoring, Frisky, Lazuli \& Whiskers) comprising 9-14 individuals, between 10th-19th November 2008. Each session commenced at 6:30 a.m., and lasted approximately 2 hours 40 mins (median 9610 seconds). 
The GPS tags acquired latitude and longitude positions with a frequency of one fix per second; further details on the GPS tagging method are provided in the Supplemental Information.

\subsection*{Using relative spatial position to identify leading individuals}
Our second null hypothesis concerns the relation between direction of information transmission and individual spatial position within the moving group.
%We first describe a technique to estimate the front and rear positions of each member of the focal pair relative to the other, which we term the relative longitudinal position, LP. 
We first describe a technique to estimate the front and rear positions of each member of the focal pair relative to the other, which we term the relative longitudinal position, LP.
Specifically, we infer from the two focal trajectories a raw ``group'' trajectory by computing the barycentric position of the focal pair of individuals. This trajectory consists of alternating quasi-static (e.g. both individuals foraging within a patch) and dynamical periods (e.g. at least one individual moving between patches). Whilst the GPS-inherent measurement error is not a problem during dynamical periods, it results in spurious scatter around the real location of the estimated focal pair barycentre during static periods, which yields erroneous trajectory estimates. In order to extract the global shape of the focal pair trajectory whilst avoiding overfitting of the noise during the static periods, we use a low-order approximation (least-squares fitted B-spline) of the pair's barycentric trajectory. The degree of the spline $k$ is chosen in each session as the minimal degree that reproduces essential components of the trajectory, such as loops and fast direction changes. As such, it varies as a 
function of the tortuosity of the recorded trajectory (with straighter paths allowing for the selection of lower degrees), with $4\leq k\leq15$ across the 6 sessions. For each time step, we then project the position of the two individuals onto the corresponding segment of the smoothed trajectory. This way, we obtain a longitudinal position estimate (termed ``Leadership index'' in Fig.~S4 and Video~S1) from the signed distance separating the individuals from their projections on the smoothed trajectory (Fig.~S3-S5, Video~S1).

\subsection*{Observables}
Until now, most applications of information-theoretic tools to infer drive-response relationships have measured dependencies within a pair of univariate time series (Wickes et al, 2007). However, as each meerkat trajectory is a bivariate time series corresponding to a sequence of paired $(\delta x_t, \delta y_t)$ displacements within the plane, for a given pair of trajectories there are several alternative observables for which the CMI may be measured.
The simplest strategy is to consider only the CMI between the two animals' movements in a single dimension, that is either the $\delta x$ or the $\delta y$ displacements. Although considering only one degree of freedom discards half of the available data, it need not necessarily discard the same in terms of the information due to correlations between the displacements in each plane \citep{wicks2007}. Hence we measured the CMI for displacements along a single axis to determine whether such a minimal approach sufficiently replicates the patterns captured by CMI analyses that utilise the paired displacements along both axes. A more comprehensive approach is to use each animals' sequence of paired $\delta x, \delta y$ displacements to calculate its displacements within the plance itself, rather than along a single axis. Here, to measure the displacement within the plane, we calculated the familiar Euclidian distance (or Euclidian norm), $\|x\|$,  and the maximum norm, $L_\infty$, also known as the Chebyshev 
distance. The coupling between these two time series could then be calculated as before.

\subsection*{Use of sliding windows to extract the CMI}
The conditional mutual information was extracted from each pair of trajectories using two sliding windows of fixed width, $w_{size}$. The first, spanning the interval $t : t+w_{size}$, was used to define a subsection of the two predictor variables, $A_t$ and $B_t$. The second window defined the movement of the putative follower at $t+lag$ time steps in the future, $B_{t+lag}$, and spanned the interval $t+lag : t+lag+w_{size}$. Then, for each of the individual increments of the first moving window (covering $A_t:A_{t+w_{size}}$ and $B_t:B_{t+w_{size}}$), the second window is repeatedly incremented forward from the current time step, $t$, up to some maximum lag in the future, $t+lag_{max}$. Thus, for a given lag the second window, defining the future response of the putative follower, covers $B_{t+lag} : B_{t+w_{size}}$. This repeated forwards-incrementing of the second window at each increment of the first allows the influence of the putative driver on the future responder follower to be measured over a range 
of temporal scales (see Fig.\ref{fig:TS_3D}. 
To best utilise the high GPS recording frequency (1~Hz), both time windows were incremented in steps of one second, $w_{step}=1$. If N represents the number of time steps in a trajectory then the incrementing process produces a matrix of $N / w_{step}$ columns representing the increments of the first window, and $lag_{max}$ rows representing the increments of the second (Fig.~\ref{fig:DIRECTIONALITY}), which we term the lag-specific CMI. To obtain a single value of directional information transmission for each time step, $t$, the mean or `lag-averaged' CMI is taken across all lags considered: 

\begin{eqnarray}
I(A \rightarrow B) = \frac{1}{lag_{max}}  \sum_{lag=1}^{lag_{max}} I(A_t;B_{t+lag}|B_t).
\end{eqnarray} 

As well as reducing the dimensionality of our main metric for measuring information flow, this averaging also serves to decrease the variance of the lag-specific CMI estimate \citep{paluvs2001,paluvs2007}.

\subsection*{Assigning statistical significance}
After obtaining some measure of directional information flow for a given pair of animals, it is desirable to assign some degree of statistical significance to the deviations from the null expectation. That is, if one is to state that any observed deviations from zero directionality are statistically significant (at a given alpha level), one must reject the null hypothesis that the observed directionality is outside the distribution of directionalities. The latter represents an ensemble of directionalities generated from pairs of null trajectories that retain the same start and end $(x,y)$ coordinates but do not exhibit any coupling.

For our null model, it is necessary to generate an ensemble of `random' trajectory realisations, against which the observed data are compared. For such a null model we generate data surrogate \citep{prichard1994,schreiber2000} trajectories. Although these random trajectories start and end at the same locations as the observed trajectories, the information transmission between pairs of such surrogate trajectories is zero by construction. Hence they represent an appropriate null model against which to test for the presence of such coupling in the observed trajectories (See Supplemental Information for a detailed description of the creation of surrogate trajectories).

Following \cite{schreiber2000}, for each trajectory pair we generated 100 surrogate trajectory pairs. Then, for each time step, we conducted a two-tailed test of the null hypothesis that the observed value of $D(A \rightarrow B)$ was drawn from the expected (surrogate) distribution; if the observed value was outside the $2.5-97.5$ percentiles of the expected distribution, the null hypothesis was rejected (at the $\alpha=$0.05 level), and the existence of drive-response roles concluded.

%%%%%%%%%%%%%%%%%%%%%%%%%%%%%%%%%%%%%%%%%%%%%%%%%%%%%%%%%%%%%%%%%%%%%%%%%%%%%%%%%%%%%%%%%%%%%%%%%%%%%%%%%%%%%%%%%
\section*{Acknowledgements}
The authors wish to acknowledge Tim Clutton-Brock for access to habituated animals and long-term records of individual life-histories, and also D. Garcia, A. Garas and T.C-B. for useful comments and discussions. We also wish to thank the Kalahari Research Trust for use of the site, and station managers Robert Sutcliffe and David Bell. 
%%%%%%%%%%%%%%%%%%%%%%%%%%%%%%%%%%%%%%%%%%%%%%%%%%%%%%%%%%%%%%%%%%%%%%%%%%%%%%%%%%%%%%%%%%%%%%%%%%%%%%%%%%%%%%%%%
\bibliographystyle{pnas}
\bibliography{Meerkats_paper}
%\printbibliography
%%%%%%%%%%%%%%%%%%%%%%%%%%%%%%%%%%%%%%%%%%%%%%%%%%%%%%%%%%%%%%%%%%%%%%%%%%%%%%%%%%%%%%%%%%%%%%%%%%%%%%%%%%%%%%%%%

\newpage

\begin{figure*}[htbp]
\begin{center}
    \includegraphics[width=\textwidth]{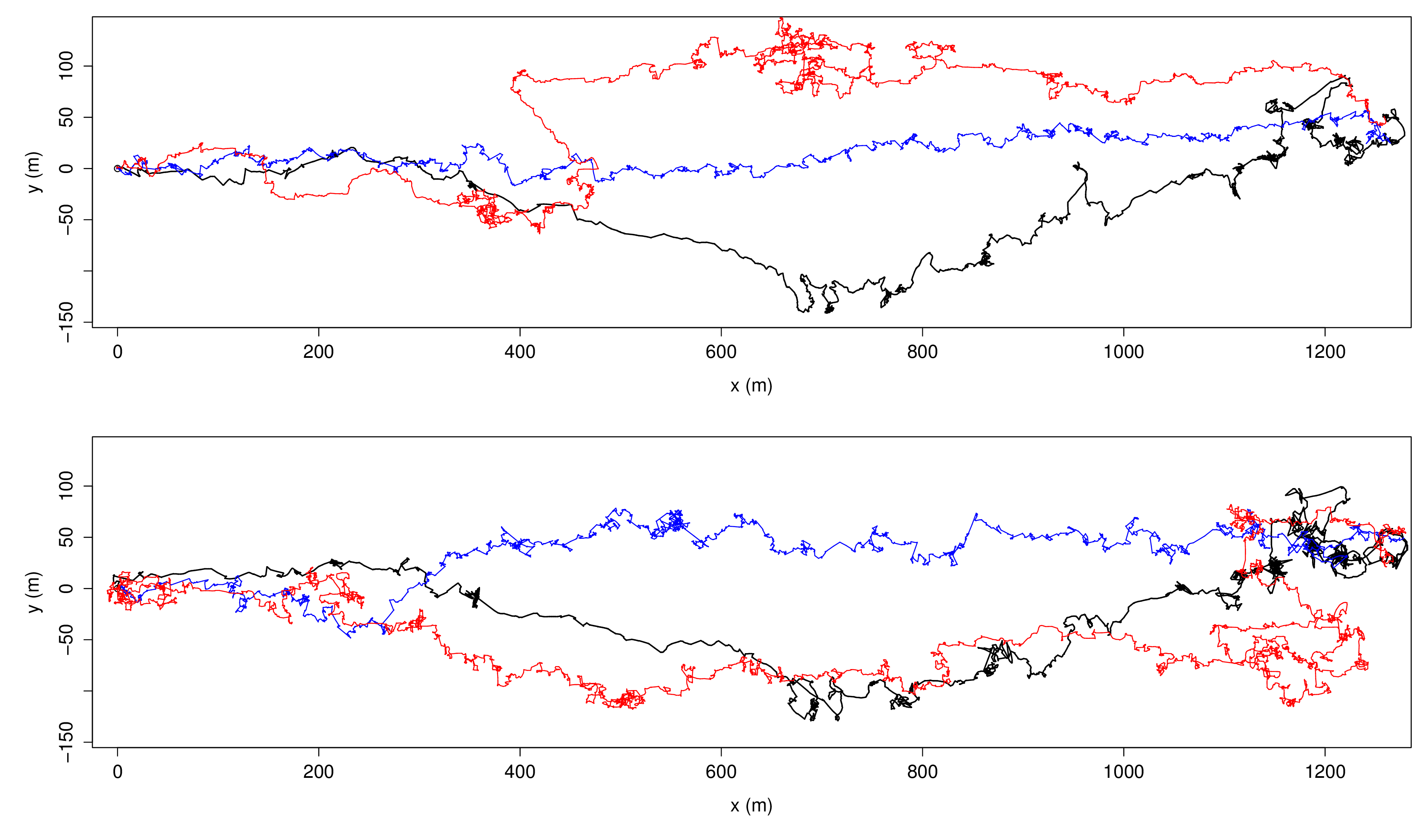}
\end{center}
\caption{Example trajectories for a pair of GPS-tagged meerkats. The black paths are the original trajectories. The surrogate trajectories are drawn in red. A simple permutation (random re-shuffling) of the sequence of (x,y) displacements is shown in blue, for comparison. For the random re-shuffling, the x and y displacements are independently shuffled. Starting and ending points are identical to all paths and the starting point is set to the origin.}
\label{fig:METHOD}
\end{figure*}

\begin{figure*}[htbp]
\begin{center}
    \includegraphics[width=10cm]{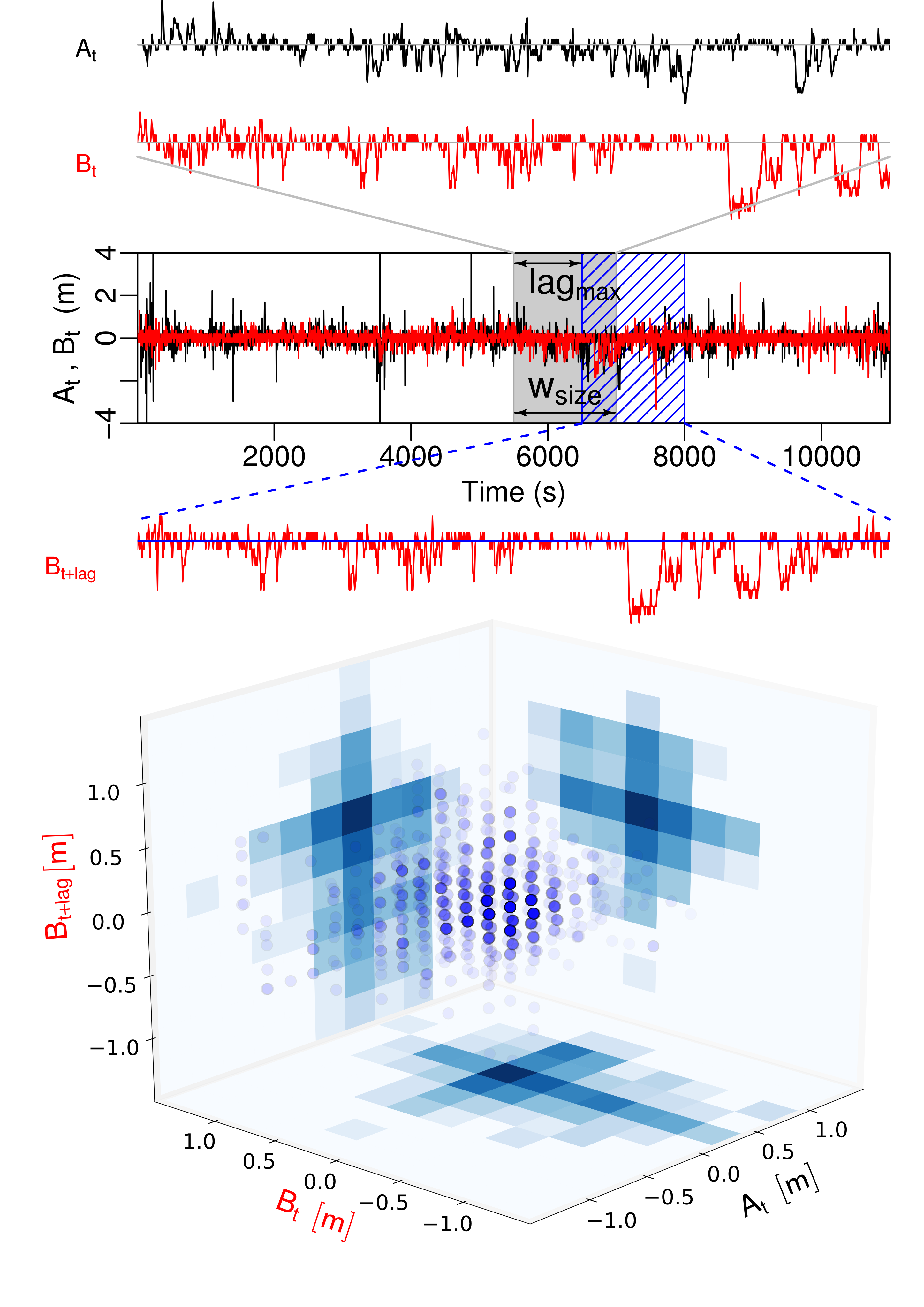}
\end{center}
\caption{Illustration of the use of time-lagged sliding windows to define the joint probability space, $P(A_t, B_t,B_{t+lag}|B_t)$, from which the CMI, $I(A_t;B_{t+lag}|B_t)$, is derived. a) The time series for the simultaneous $x$ displacements of two individuals, A and B (Session 13, total duration 3 hours). Magnified time series from the reference window (grey rectangle) and the time-lagged window (blue-hashed rectangle) are shown immediately above and below respectively. The upper two time series give the putative driver, $A_t$, and the conditional variable, $B_t$. The lower time series gives the lagged response, $B_{t+lag}$. b) Three-dimensional representation of $A_t$, $B_t$ and  $B_{t+lag}$ for the grey \& blue rectangles in panel a. The shaded grids on the floor and on each wall are two-dimensional histograms depicting the joint distributions used in the calculation of the CMI. For example, the MI of the joint distribution on the right wall gives the self-information of animal B, namely $P(B_{t+lag}
|B_{t})$ (the extent to which its current actions influence its future actions). The MI of the joint distribution on the left wall gives the influence of the animal A on the future of B, $I(A_{t};B_{t+lag})$, which includes the possible influence of B upon itself).}
\label{fig:TS_3D}
\end{figure*}

\begin{figure*}[htbp]
\begin{center}
    \includegraphics[width=\textwidth]{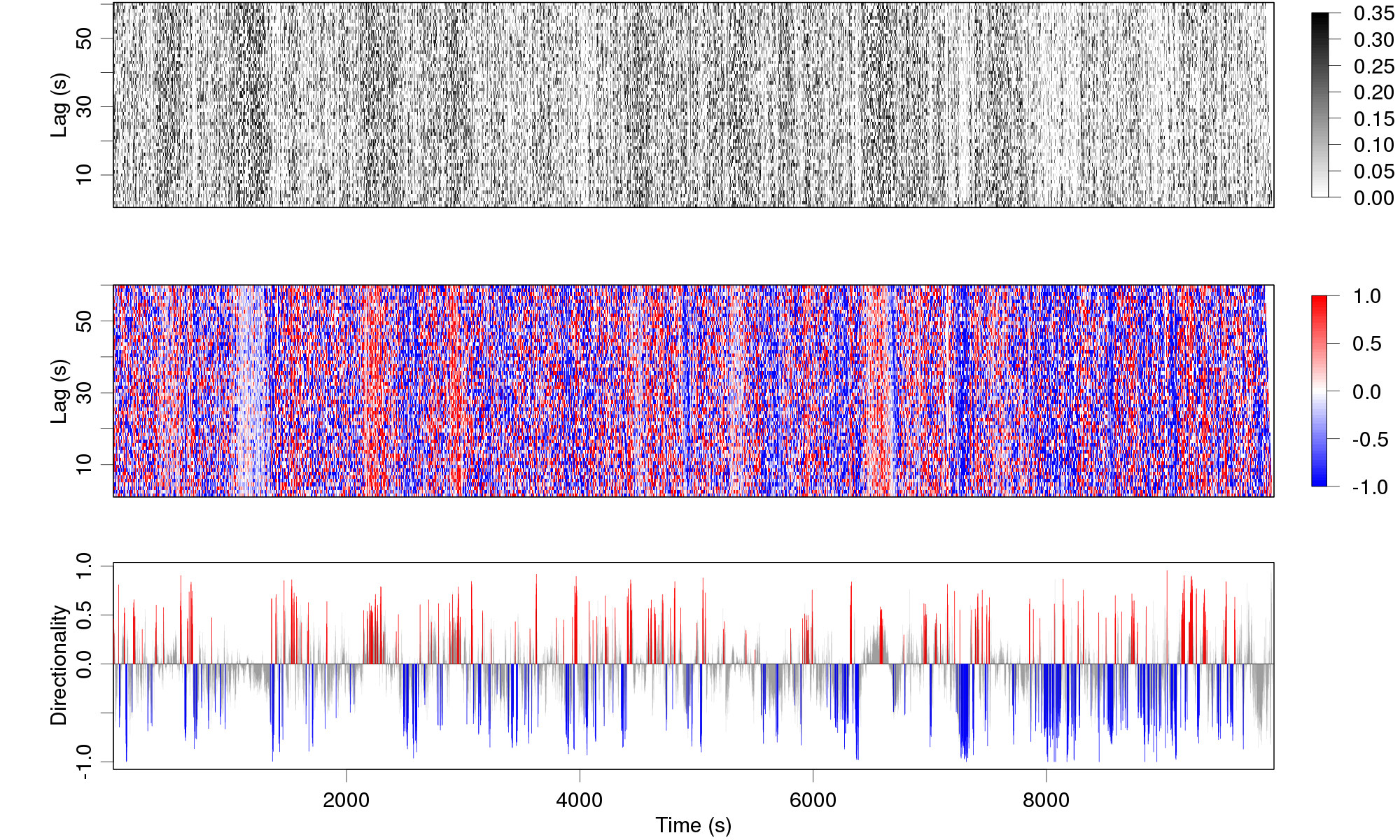}
\end{center}
\caption{Illustration of the steps leading towards the calculation of the directionality. a) The lag-specific directional information flow, $I(A_t;B_{t+lag}|B_t)$, as a function of time (the $x$ axis is common to all plots). The information flow in the opposite direction, $I(B_t;A_{t+lag}|A_t)$, is not shown for clarity. The side ribbon gives the CMI in nats. b) The lag-specific directionality, calculated directly from  $I(A_t;B_{t+lag}|B_t)$ and $I(B_t;A_{t+lag}|A_t)$. The side ribbon gives the (unitless) directionality. Positive values indicate that animal A influences B, negative that B influences A. c) The lag-aggregated directionality, $D(A \rightarrow B)$, calculated from the lag-averaged CMIs. Time steps in which the observed directionality was either significantly greater or less than the expected value (from the surrogate trajectory ensembles), are indicated in red or blue respectively. Grey line segments indicate no significant difference. All plots have $w_{size}=15~\hbox{s}$, and represent the 
CMI measured on the Euclidian distance, $\|x\|$.}
\label{fig:DIRECTIONALITY}
\end{figure*}

\begin{figure*}[htbp]
\begin{center}
    \includegraphics[width=6cm]{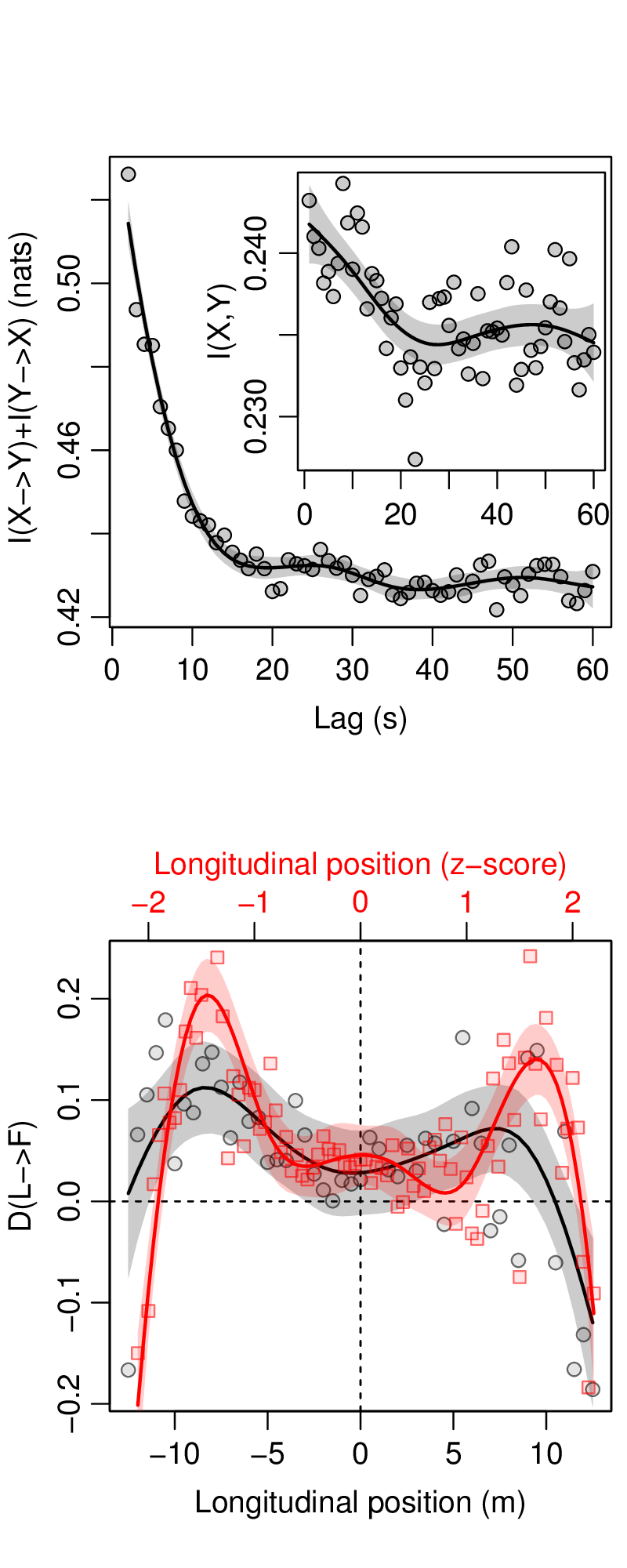}
\end{center}
\caption{Directional and absolute measures of information flow in time and space. 
a) The net two-way information flow magnitude, calculated from the sum of the two directional CMIs, $I(A \rightarrow B) + I(B \rightarrow A)$, as a function of the time lag. The panel inset shows the mutual information, $I(A_t,B_{t+lag})$.
b) Directional information flow, as measured by the directionality from the driving individual (L) to the responding individual (F), $D(L \rightarrow F)$, as a function of the longitudinal position of the driver. For each pair, the animal with the greatest mean directionality over the entire session was defined as the driver. Each point represents the mean of the six drivers (each from a different recording session). 
Black line: directionality as a function of the absolute longitudinal position. Red line: directionality as a function of the standardised longitudinal position. 
Lines and shaded areas represent spline-smoothed fits (p-splines) and 95\% confidence intervals. The horizontal blue lines give the mean and 95\% CI for the expected directionality, $D(L_{surr} \rightarrow F_{surr})$, for the surrogate trajectories.
All panels show the information flow calculated on the Euclidian distance, $\|x\|$, and used the same parameter combinations, namely, $w_{size}=15~\hbox{s}$, and $lag_{max}=60~\hbox{s}$.}
\label{fig:FLOW_VS_FB_DIST}
\end{figure*}

\begin{table}
\centering
\begin{tabular}{lllllll}
$w_{size}$& $x,y$ (\%)&$\|x\|$ (\%)&$L_\infty$ (\%)\\
\hline
5  & 2.7  & 4.6  &2.8  \\
10 & 5.6  & 7.9  &5.7  \\
15 & 8.9  & 13.1 &9.7  \\
25 & 12.7 & 19.5 &15.6 \\
\end{tabular}
\caption{The proportion of time steps in which two-tailed randomisation tests detected statistically significant directionality at the $\alpha=0.05$ significance level, as a function of the moving window width $w_{size}$. Figures represent the mean taken across all six sessions, for combined $x$ and $y$ displacements ($x,y$), Euclidean distance ($\|x\|$) and Chebyshev distance ($L_\infty$).}
\label{tab:PERCENT_SIGNIF}
\end{table}

%%%Supplemental Information
\newpage


\section*{Supplementary material}

# Supplemental Information:
# Dynamical coupling during collective animal motion

T.O. Richardson, N. Perony, C.J. Tessone,

C.A.H. Bousquet, M.B. Manser, F. Schweitzer

# Contents

1

## S.1 GPS tagging

We used GPS tracking devices (TechnoSmArt, Rome, Italy) embedding LEA-4 and LEA-5 chips (u-blox AG, Thalwil, Switzerland). These devices were manually attached to collars on the animals at the start of the observation period with a velcro tape so that handling of the animals was minimised. The weight of the total collar, with battery, protection and additional radio beacon (BD-2, Holohil Systems Ltd.) to make it fit for field deployment was kept under 20 g, which corresponds to less than $4\%$ of the weight of an adult meerkat (600 to 900g). The data was dumped to raw logs, then converted to the NMEA format with the GiPSy software (TechnoSmArt, Rome, Italy, http://www.technosmart.eu/gipsy.php) and parsed with MATLAB (The MathWorks, Inc.). Figure S1 shows the GPS device and a picture of a male adult meerkat wearing it.

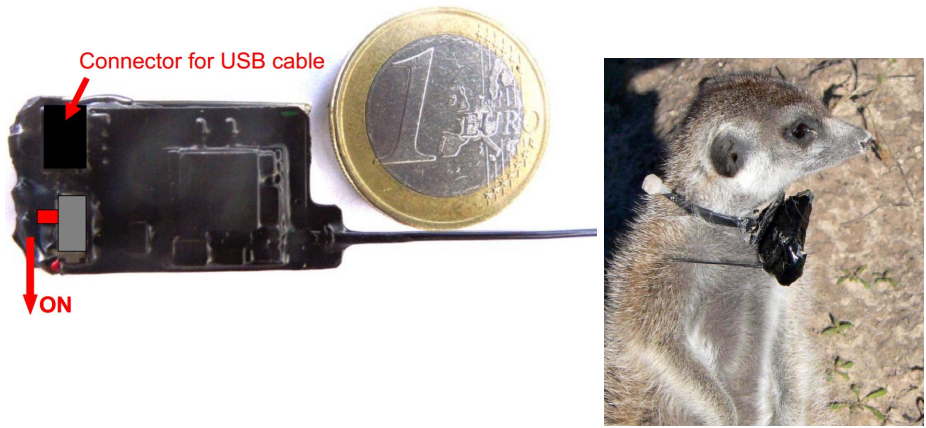

Figure S1: **(left)** GPS module (top side). The coin is pictured for size comparison (picture courtesy of TechnoSmArt). **(right)** Adult male meerkat wearing the module as a collar, with the antenna of the radio beacon wrapped around the collar and the GPS antenna protruding from the GPS module.

## S.2 Using asymmetrical information flow to measure the direction and magnitude of leadership

In this section we will describe in more detail the information-theoretic components of our framework that allow for the extraction of both the direction and strength of pairwise information transmission. Ideally, the input data will be simultaneously-recorded paired trajectories in which the inter-location intervals are fixed, and in which the location-acquisition

2

times are the same for both animals. Departures from these ideal requirements are acceptable if the deviations represent stationary noise, for example occasional failures to acquire GPS signals. However, when the input data consistently deviate from the regularly sampled ideal, artefacts may be introduced into the assessment of information flow. For example, commonly-used techniques for preserving GPS battery life, such as powering-down the device for 40 seconds in each minute, can introduce both periodicity, and drift due to accumulated variation time taken to turn the device on or off.

If the relationship between the purported driver and response variables is thought to be non-linear, then non-parametric correlation metrics (e.g. Spearman's $\rho$) are often preferred due to their ability to capture non-linear relationships. However, if for each value of the predictor there exists more than one response, even non-parametric correlations fail (Fig. S2). We therefore advocate an information-theoretic approach, as rather than assuming that the driver and response are related by some underlying function in which each value of the predictor gives a single response (i.e. a bijective function), approaches based on mutual information instead quantify the reduction in uncertainty about one variable obtained by measuring the other by comparing the joint (bivariate) distribution of predictor and response with the two marginal distributions.

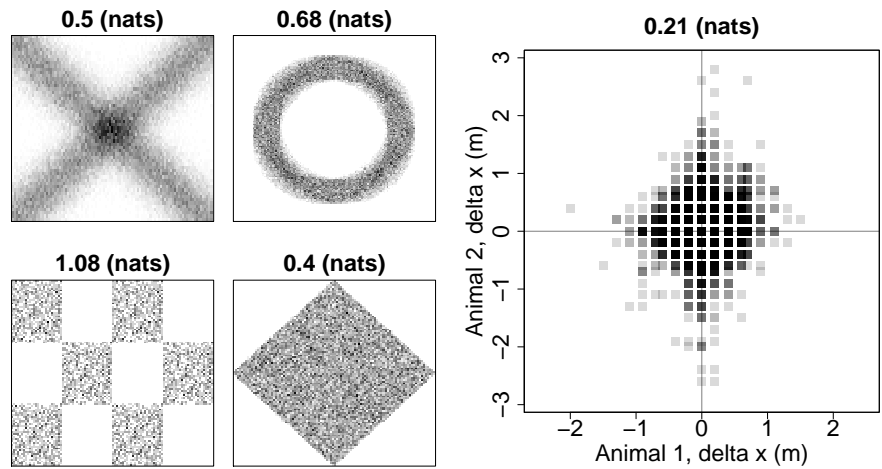

Figure S2: Patterns of non-random association that both parametric (Pearson's $r$) and non-parametric rank-based measures (Kendall's $\tau$ & Spearman's $\rho$) fail to capture. a-d) All three correlation metrics give zero correlation for these four associations. e) Observed association between the displacements along the $x$-plane for two animals. The mutual information (in nats) between the x and y axes, $I(A; B)$, is indicated above each plot.

In order to determine if a given member of the pair is the leader (and hence also, if the other member is the follower) we examine the time-ordering of its instantaneous displacements

3

in either a single plane (i.e. the $x$ or $y$ dimensions), or its instantaneous displacement (e.g. Euclidian distance) within the plane itself. We then determine by how much knowledge of the putative leader's movements reduces the uncertainty about the subsequent movements of the putative follower.

Formally, the mutual information between two potentially-related variables is the reduction in uncertainty (i.e. the entropy decrease) in the first variable, A, given prior knowledge of the other, B. The MI can either be expressed in terms of the marginal and joint entropies (as below), or in terms of its marginal and joint probability distributions. However, for simplicity, we present the former,

$$I(A; B) = H(A) + H(B) - H(A, B) \tag{S1}$$

The calculation of the Shannon entropy of a distribution, (e.g. $H(B)$) requires the use of a logarithm [1]. The unit of information is termed a 'bit' when the base in the logarithm is 2, and when the natural log is used -as here- the units are termed 'nats'.

For the purposes of inferring coupling between pairs of animal trajectories, we substitute the current state of the putative leader, $A_t$, and the future state of the putative follower, $B_{t+lag}$ into the above. Then, the MI, $I(A_t; B_{t+lag})$, quantifies the reduction in the uncertainty about the future of $B$, gained from knowing the current state of $A$.

However, a large value of the $I(A_t; B_{t+lag}) > 0$ should not be taken to infer that A leads B, because in many real-world time-series there is a non-zero mutual information between both the current and future states of the putative follower itself, that is $I(B_t; B_{t+lag}) > 0$ [2]. Therefore, to access the net information transfer from A to B, one must remove the possible influence of B upon itself in the future. To do so we turn to the Conditional Mutual Information (CMI), which is defined as the reduction in uncertainty about the future state of the hypothesised follower, $B_t$, derived from knowledge of both its own current state, $B_t$, and that of the hypothesised leader, $A_t$.

$$I(A_t; B_{t+lag}|B_t) = H(A_t, B_t) + H(B_{t+lag}, B_t) - H(A_t, B_{t+lag}, B_t) - H(B_t) \tag{S2}$$

A non-zero value of the $I(A_t; B_{t+lag}|B_t)$ can be interpreted as an influence of A upon the future of B, controlling for the influence of B upon itself [3].

### S.2.1 Practical estimation of MI, CMI & directionality

As with measures of correlation, information-theoretic measures of entropy, MI or CMI, are estimations of the underlying order shared between the specified variables, and as such

there are many different computational methods availabe for their quantification. A major problem with many MI estimators is that for small sample-sizes where the joint probability space, $P(x,y)$, appears more granular, the estimation is positively biased [4]. The problem is not merely theoretical; GPS data often contain missing gaps which may be irregularly distributed throughout the trajectory, hence the sliding-window method for measuring the CMI dynamics will often include considerable fluctuations in the sample size. To reduce the problem of bias, we used the K-NN method [5] which estimates the mutual information between two variables $x, y$ by measuring the distribution of distances from each point in the $P(x,y)$ space to its $k^{th}$ nearest-neighbour. When the neighbour is the second nearest ($k=2$), the method exhibits almost no bias at either small or large sample sizes [2], hence we set $k=2$.

Whilst the CMI measures the flow of information, $I(A_t; B_{t+lag}|B_t)$, from the animal in the first term, $A_t$ to the future state of its partner, $B_{t+lag}$, this is not necessarily the converse of the flow of information in the opposite direction, namely, $I(B_t; A_{t+lag}|A_t)$. Hence, for a pair of individuals, it is possible to find non-zero information-flow in both directions. Thus, we calculate the aggregate of the two, which is termed the directionality [6].

$$D(A \to B) = \frac{I(A \to B) - I(B \to A)}{I(A \to B) + I(B \to A)} \tag{S3}$$

As a ratio of bi-directional information flow, the directionality condenses the asymmetry of information flow into a single dimensionless number. The directionality is a symmetric function, so if the animal A leads B, then $D(A \to B)$ will be positive, and by definition, $D(B \to A)$ will be negative with equal magnitude.

## S.3 Measuring the spatial position within the pair

### S.3.1 Measuring the relative distance of each member to the pair's center

In the main text, we describe the procedure used to measure the distance of each group member to the barycenter of the group, which we term longitudinal position. Whilst we apply this technique to pairs of individuals only, the method is suitable to be used with groups of any size. In fact, its accuracy increases with the size of the group, as fluctuations of the group's "raw" trajectory due to finite-size effects (i.e. the sporadic movement of one individual influencing the trajectory of the whole group's barycenter) are diminished. Fig. S3 illustrates the technique for one GPS session with two individuals (sampling rate

1 Hz, 11'036 fixes, 184 minutes).

Figure S3 shows that whilst the individual paths (thin coloured curves) are very tortuous and alternate between static and dynamic periods, the spline approximation of the pair's trajectory (thick black curve) conserves only the most significant features, namely the general orientation, reversal of foraging direction (twice), and a loop (radius $\approx 20$ m, orbit time $\approx 40$ minutes). The inset shows the projection of the individual positions at a given time on the spline approximation, which is used to infer the longitudinal position of each individual within the pair. Additionally, Video S1 shows an animation of this technique applied to the calculation of the longitudinal positions of two individuals in two different instances.

### S.3.2 Standardising longitudinal distances

There was considerable between-session variation in the distributions of the relative longitudinal distance, LD (Fig. S5). This variation might be due to variation in the physical or biotic structure of the environment through which each foraging group moved, may have contributed to the high between-pair variation in the range of the LD distributions. For example, spatial patchiness of prey items might cause the progression of the collective trajectory to become intermittent. Spatially clustered distributions of physical or visual obstacles could impede movement or vision, which would likely have similar unforeseen impacts upon the LD range.

Irrespective of the origin of the LD variation, a naive calculation of the mean information flow, $D(L \to F)$, as a function of the LD is adversely affected by the varying cardinality of the number of observations in each bin. This effect becomes problematic at extreme LD values; for example, only two sessions contributed to the LD bins above 18m. One solution - which we adopt - is to remove LD distances above a threshold beyond which too few sessions populate each bin (here the threshold was +/- 12.5 metres). We therefore present an additional standardisation of LD that can be used to average the directionalities across pairs with widely-varying LD distribution ranges. To do so, for each session we convert the LDs to z-scores by subtracting the mean and dividing by the standard deviation, where the mean and standard deviation are specific to the particular session.

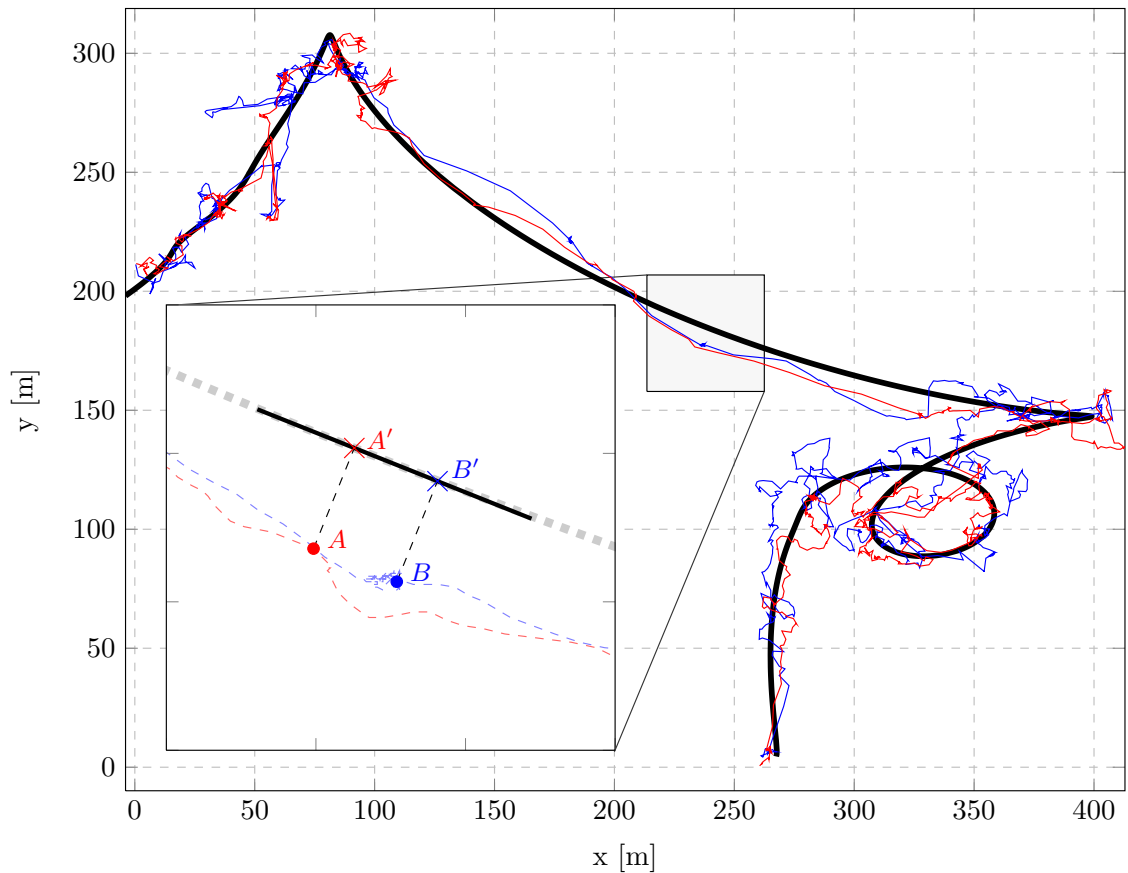

Figure S3: Individual paths (thin coloured curves) and B-spline approximation of the pair's center's trajectory (thick black curve) for one data recording session. In the inset (magnified section of the main plot), A and B correspond to the position of both individuals at a given time. A' and B' are their respective orthogonal projections on the segment (thick black line) tangent to the smoothed trajectory (dashed grey curve) at this point in time. Because this segment is oriented (see also Video S1 for the dynamics of this trajectory), it is possible to calculate a signed distance from each individual's position to the pair's center. In this case, individual A is 3 metres ahead of the pair's center and B is 3 metres behind (i.e. A is 6 metres ahead of B).

7

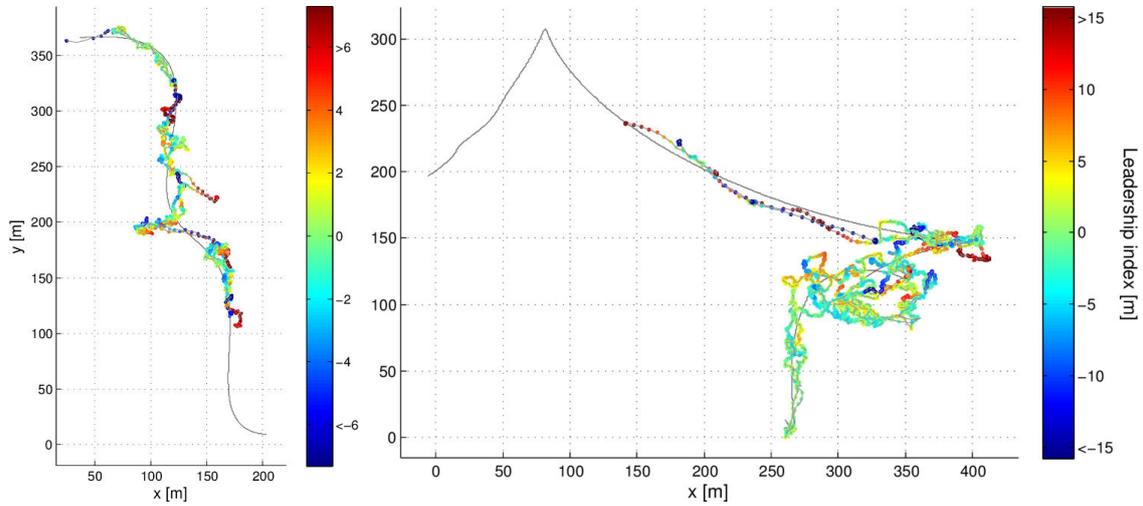

Figure S4: / Video S1 (video at `http://youtu.be/exl-1KpsONM`): animated trajectories of two individual meerkats in two different recording sessions (both sessions recorded at 1 Hz. **(Left)**: 7'714 GPS fixes, 129 minutes. **(Right)**: 11'036 fixes, 184 minutes). The individual paths are overlaid on top of the group's smoothed trajectory. The colour of the dots (paths sampled every 5 seconds) indicate the corresponding distance from the pair's barycenter, or signed "leadership index". 1 second in the video corresponds to 5 minutes of data for both sessions.

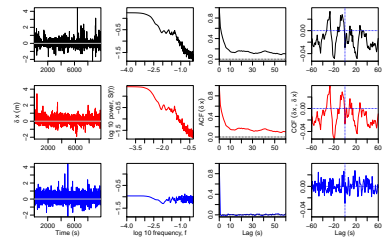

Figure S5: The between-session variation in the range of the relative longitudinal position, LP. **(Top)**: The observed LP for each session. The dashed red lines indicate the arbitrary cutoff (+/- 12.5m). **(Bottom)**: The LP, standardised such that the units are z-scores (units of the session-specific standard deviation). The red lines indicate the non-arbitrary cutoff, defined by the smallest standardised distance at which all sessions are represented (z=-2.14 & z=2.18).

8

# S.4  Creating surrogate trajectories

Surrogate time-series methods were originally developed to identify non-linearity within *single* (i.e. univariate) time-series [7], although they have also been applied to the study of drive-response relationships *between* pairs of univariate time-series, for example between the human heart and breathing rates [8] or between electroencephalograph recordings from different brain regions [9].

When applied to time-series data, simple shuffling-based randomisations destroy many of the temporal correlations present in the original, whereas data surrogates are 'constrained realisations' of the original time-series [10, 11] that preserve many of its measurable temporal properties (Fig. S6).

Multivariate surrogates which preserve both the temporal correlations within each time-series (e.g. the autocorrelation function, power spectrum), whilst also preserving the correlations between the different time-series (e.g. the cross-correlation and coherence) are also feasible [8, 10]. However, the trajectories of GPS-tagged animals are actually best described as a bivariate time-series, consisting of two simultaneously-recorded sequences of displacements along each axis, $(\delta x_t, \delta y_t)$ (Fig. S6).

Hence, it is desirable that each surrogate trajectory should retain the within-trajectory statistical characteristics of the original whose within-trajectory statistics it replicates, whilst entirely removing the between-trajectory correlations hypothesed to exist between the members of the pair. To produce the surrogate data, we used an iterative Fourier-based procedure [7], implemented in the freely-available TISEAN (version 3.0.1) software package [12]. The preservation of this within-trajectory cross-correlation between sequential pairs of $(\delta x, \delta y)$ displacements is important because they are negatively correlated (Fig. S6); the upper mechanical limit on the instantaneous speed, means that a large displacement along one axis necessarily leads to lower displacements along the second axis. Were the two components of each trajectory - the $(\delta x, \delta y)$ displacements - to be treated as independent time-series, and univariate surrogates produced for each, the resulting trajectory surrogate would preserve the temporal correlations for movement within the $x$ and $y$ planes, but it would not retain the negative cross-correlation between the displacements along the two axes, resulting in an overrepresentation of large Euclidian displacements.

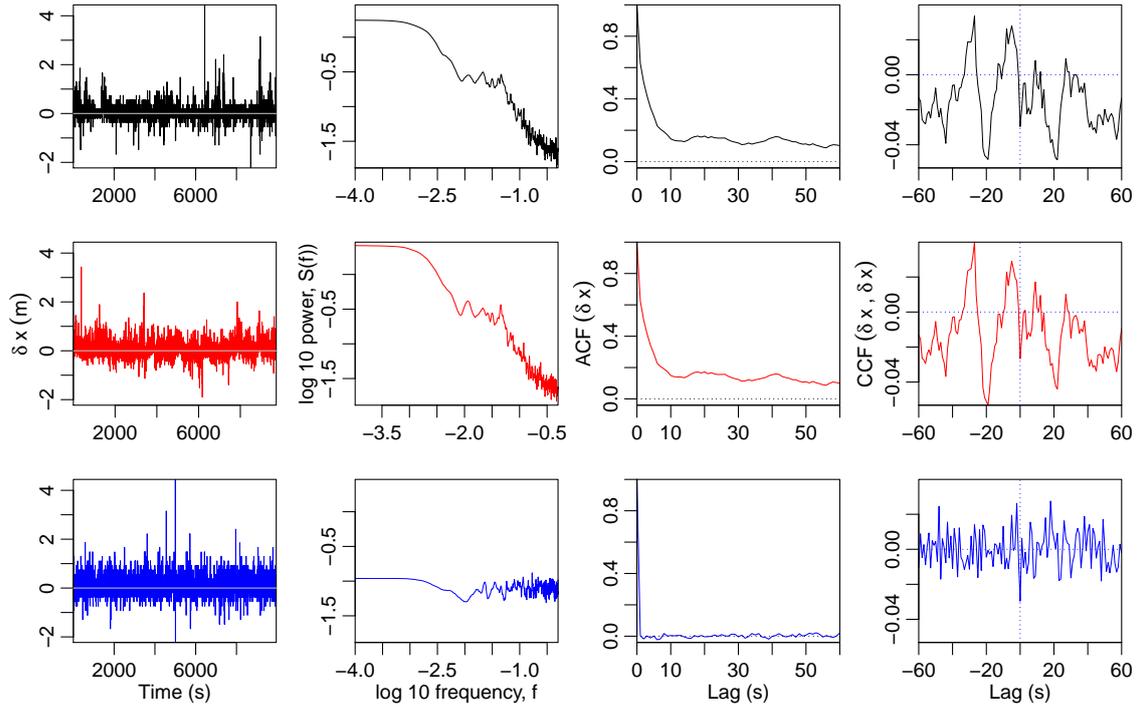

Figure S6: Illustrating temporal statistics of a single empirical trajectory (black), a surrogate path (red), and a reshuffled path (blue). The statistics presented here are for the trajectory shown in Fig. 3 in the main paper (colours herein correspond the the colours used there). **(Left)**: The univariate time-series of displacements along the x-axis. **(Left-middle)**: Double-log plot of the power spectrum for the $\delta x$ time-series. **(Right-middle)**: The autocorrelation function (ACF) for the $\delta x$ time-series. **(Right)**: Cross-correlation function (CCF) for the bivariate time series composed of paired displacements along the x and y axes. The surogate trajectory preserves both the univariate and bivariate temporal correlations, whereas the simple randomisation preserves neither.

## S.5 Temporal statistics of directionality

In this section we explore the temporal characteristics of the fluctuation in the directionality, $D(A \to B)$. We use the autocorrelation function to measure the self-correlation of $D(A \to B)$. The decay of the ACF exibits memory effects up to approximately 40 seconds delay. Similarly, when we consider the spectral density tail, we observe an exponent between -1 and -2, which indicates the presence of long-term temporal correlations. These temporal statistics were higly consistent across all six sessions.

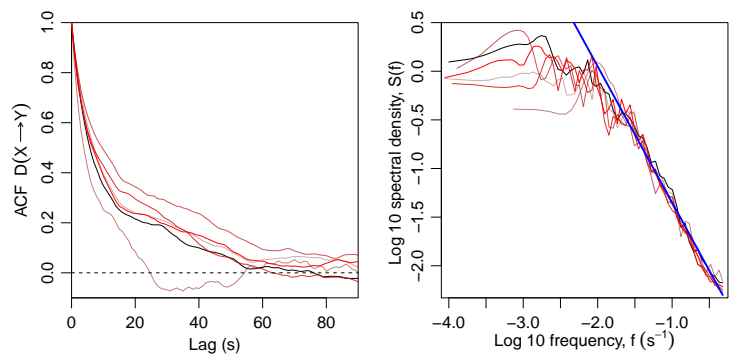

Figure S7: Characterising the temporal correlations in the directionality time-series. **(Left)**: The autocorrelation function (ACF) for all six directionality time-series. **(Right)**: The power spectrum for all six time-series. The blue line is a guide to the eye, with slope -1.4. Both panels give the directionality measured on the Euclidian displacements, $\|x\|$. Almost identical statistical characteristics were found for the directionality measured on and the $\delta x$, $\delta y$ and $L_\infty$ displacements.

## S.6 Selection of time-lag and sliding-window size

When measuring MI as $I(A_t, B_{t+lag})$, or CMI as $I(A_t; B_{t+lag}|B_t)$, the choice of both the time-lag and the sliding window width, $w_{size}$, are free parameters. These choices should be informed not only from expectations derived from consideration of the biology of the study system (e.g. cognitive ability, typical activity rates), but also by studying the effect of systematically varying their values. Below we outline the effect of systematically varying $w_{size}$ upon the decay of the CMI as a function of the time-lag (Fig. S8).

The choice of the window size, $w_{size}$, influences the measured values of both the MI and the CMI; too narrow and noise dominates any underlying signal, too large and the alterna-

11

tion of pairwise leader-follower roles occurring on short time-scales may be lost. Similar considerations apply to the choice of a maximum time-lag; too-short lags may not allow sufficient time for the putative follower to react to the actions of the leader.

The effect of a too-narrow sliding window or a too-short time-lag becomes relevant if indirect interactions can act as alternative routes for information propagation. For example in the indirect interaction $A \to B \to C \to D$, each step in the chain would increase the time-lag required to capture the causal link $A \to D$. However, with each additional link, the net amount of information transmitted from A to D would probably be attenuated, so it is likely that beyond intermediate temporal scales the influence of such higher-order interactions fades out. Similarly, time-lags greatly exceeding typical direct and indirect action-reaction times may incorporate unwanted noise from externalities, such as the putative follower reacting to the myriad of extrinsic events which are not contained in the GPS records.

To choose appropriate values of the sliding window width we systematically varied $w_{size}$ within the range (5-40 seconds) and examined the CMI as a function of the time-lag. The functional form of these decay curves varies in a consistent manner according to the window size (Fig. S8). Narrow windows ($\leq 15$ sec) show an extremely steep decay of the coupling as a function of the time-delay, which is successively lost for wider windows. In terms of the time-lag, all window sizes agreed that the decay reaches a baseline after approximately 30-60 seconds. For the analysis presented in the main paper we therefore used $w_{size}$=15 seconds, and $lag_{max} = 60$ s.

## S.7 Damping GPS noise

The precision of GPS tracking is affected by weather conditions, the degree of environmental 'clutter' such as patchy vegetation, and also in this case by short periods in which the foraging individual was digging and so slightly below ground-level. Hence a first step was to reduce the influence of the spurious points associated with GPS noise.

To damp such fluctuations, a fixed-width sliding window was incremented through the trajectory, partitioning it into a sequence of overlapping spatial point-patterns. The sliding window had width $w$ (where $w$ is an odd number), so among the set of $w$ points, the focal point (recorded at time $t$), is bookended by an equal number $(w-1)/2$ of preceeding and trailing points. Each of the successive point patterns was bounded by a rectangular window, defined by the minimum and maximum $x, y$ coordinates, the dimensions of which varied from time-step to time-step as the speed of the animal fluctuates. The spatial intensity function was then calculated for each successive point pattern, using Gaussian kernel density

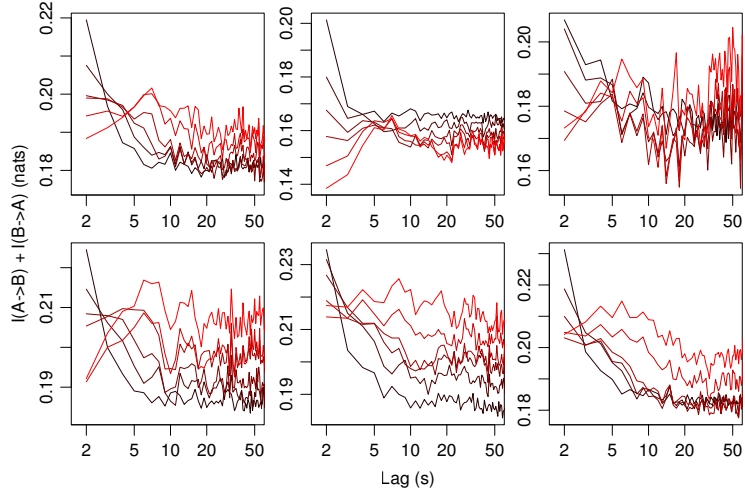

Figure S8: Exploring how varying the sliding window width, $w_{size}$, influences the relationship between CMI and time-lag. Note the x-axis is logged. Each panel represents a different GPS session. Line colours give $w_{size}$, with values of 5 (black), 10, 15, 20, 30 and 40 (red) seconds.

estimation [13], and the $x, y$ coordinate at the peak of the intensity surface (the point of maximum intensity) was then defined as the new (smoothed) coordinate.

At its simplest, a point pattern for which an intensity surface is produced using Gaussian kernel smoothing assumes all of the points within the spatial window have the same importance, and thus assigns them the same weight. However, the objective of path smoothing using a moving window was to use locational information from coordinates that were temporally proximate (though not necessarily spatially so) to the focal point in order to increase the confidence associated with the true location of the focal point. Therefore those points closer in time (but not necessarily closer in space) to the focal point at time $t$ contain more information about the true location of the focal point itself than do the more temporally distant points. Hence the weights assigned to each point were an inverse function of their delay (either forwards or backwards in time) from the focal point. More specifically, the point weight was set to decay according to a Gaussian function centred on the focal point (with mean=0), thus the smaller the standard deviation of this Gaussian function (henceforth referred to as the time bandwidth), $\sigma_t$, the smaller the influence of the temporally distant points upon both the shape of the intensity surfae and the location of its peak.

As the inter-point spacing varied over successive point patterns, using a fixed spatial bandwidth for all point patterns would result in over-smoothing for point patterns in which the points are widely-separated, and under-smoothing for the reverse case [13, 14]. There-

13

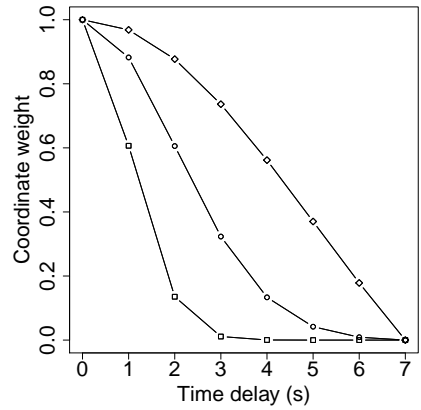

Figure S9: Weighting the $x, y$ coordinates within each the smoothing window by the inverse of the time separation from the current focal point, $t$, which by definition has time delay=0 sec. The weights are normalised so that the coordinate at both the extreme left and right ends of the moving window, have zero weight, and the focal coordinate has a weight of one. Squares; $\sigma_t$=1 sec, circles; $\sigma_t$=2 sec, diamonds; $\sigma_t$=5 sec.

fore, rather than choosing an arbitrary smoothing the spatial bandwidth was chosen using adaptive data-based bandwidth-selection, in which the bandwidth depended upon the distribution of inter-point distances. However, for a given spatial point pattern, the same spatial 'bandwidth', $\sigma_s$, (the standard deviation of the Gaussian kernel), was used for both the $x$ and $y$ dimensions, hence the kernel-smoothing was isotropic.



\end{document}